\documentclass[letterpaper, 10 pt, conference]{ieeeconf}  
\IEEEoverridecommandlockouts                              
\usepackage{graphicx}
\graphicspath{{figures/}}
\usepackage{amsmath} 
\usepackage{amssymb}  
\usepackage{cleveref}
\crefname{equation}{}{} 
\Crefname{equation}{Equation}{Equations} 
\Crefname{figure}{Fig.}{Figs.} 

\DeclareMathOperator*{\argmax}{argmax}
\DeclareMathOperator*{\argmin}{argmin}

\usepackage{flushend}

\title{\LARGE \bf
Local-Global Learning of Interpretable Control Policies: The Interface between MPC and Reinforcement Learning
}

\author{Thomas Banker\textsuperscript{\textdagger}, Nathan P. Lawrence\textsuperscript{\textdagger}, and Ali Mesbah
\thanks{The authors are with the 
        Department of Chemical and Biomolecular Engineering, 
        University of California, Berkeley, CA 94720, USA.
        {\tt\small input@nplawrence.com, mesbah@berkeley.edu}}%
        \thanks{This work was supported in part by the U.S. National Science
Foundation under grant 2317629 and the U.S. Department of Energy, Office of Science, Office of Fusion Energy Sciences under award DE‐SC0024472.}
        \thanks{\textsuperscript{\textdagger}These authors contributed equally to this work.}
}

\begin{document}

\maketitle
\thispagestyle{empty}
\pagestyle{empty}

\begin{abstract}
%

Making optimal decisions under uncertainty is a shared problem among distinct fields.
While optimal control is commonly studied in the framework of dynamic programming, it is approached with differing perspectives of the Bellman optimality condition.
In one perspective, the Bellman equation is used to derive a global optimality condition useful for iterative learning of control policies through interactions with an environment.
Alternatively, the Bellman equation is also widely adopted to derive tractable optimization-based control policies that satisfy a local notion of optimality.
By leveraging ideas from the two perspectives, we present a local-global paradigm for optimal control suited for learning interpretable local decision makers that approximately satisfy the global Bellman equation.
The benefits and practical complications in local-global learning are discussed.
These aspects are exemplified through case studies, which give an overview of two distinct strategies for unifying reinforcement learning and model predictive control.
We discuss the challenges and trade-offs in these local-global strategies, towards highlighting future research opportunities for safe and optimal decision-making under uncertainty. 
\end{abstract}


\section{Introduction}

Dynamic programming offers a theoretical framework for studying sequential decision-making problems \cite{bellman_dp_1957}.
Ensuring optimality of online decisions is paramount for the economic value, logistical efficiency, and safety of autonomous systems.
As such, research communities spanning control theory and artificial intelligence have contributed to this class of problems with different motivations \cite{sutton_reinforcement_2018, powell2011ApproximateDynamic, bertsekas_neuro-dynamic_1996}.
A core insight is the Hamilton-Jacobi-Bellman equation \cite{kalman1963theory}, or simply Bellman equation nowadays \cite{sutton_reinforcement_2018}, an optimality equation describing the dynamic relationship between the optimal present ``value'' of a decision to that of future decisions.
However, it is well-known that the Bellman equation generally cannot be solved explicitly, broadly due to the inherent uncertainty and scale of most practical problems \cite{bertsekas2012dynamic}.
Thus, the varying motivations across research communities, in combination with the set of challenges embedded in the Bellman equation, have led to vastly different perspectives for solving the dynamic programming problem \cite{bertsekas2022lessons}.

One perspective emphasizes that the Bellman equation is a \emph{global} optimality condition.
Under this view, iterative schemes are derived based on dynamic programming theory to achieve increasingly precise approximations to the Bellman equation across the entire state space \cite{bertsekas_neuro-dynamic_1996, bertsekas2012dynamic}.
Such approximations would then, theoretically, readily enable optimal decisions in any situation.
This stems from a \emph{local} perspective of the Bellman equation, which posits that the agent greedily maximizes its value function at each time step.
The local view lends itself to designing tractable optimization-based control policies that satisfy a local notion of optimality in the dynamic programming sense \cite{rawlings2017model}.

The perspective one takes towards the Bellman equation influences the approximators employed to solve it.
The global perspective is common in reinforcement learning (RL) \cite{sutton_reinforcement_2018,schulman2017ProximalPolicy}.
In the RL setting, practical algorithms should design agents that accumulate a wide range of experience, thereby aligning the agent's representation of value with the theoretical Bellman equation.
This then necessitates fairly general-purpose function approximators, such as deep neural networks, to faithfully capture the Bellman condition over a rich set of training samples.
On the other hand, optimization-based control is emblematic of the local perspective.
A widely successful example is model predictive control (MPC) \cite{rawlings2017model,lee2011Modelpredictive}, which incorporates dynamic models, operational requirements in the form of input or state constraints, and control objective functions into its design.
Consequently, optimization-based control strategies facilitate a local sense of value in an interpretable structure.
A local-global perspective should then account for the intricacies of local decision makers and the demands of global optimality conditions.


The aim of this paper is to delineate the local-global perspective underlying dynamic programming; in particular, the role of local optimization-based approximators in a global Bellman equation framework.
To this end, we discuss ways in which largely distinct research areas of RL and MPC can work in a unified fashion towards designing tractable local decision makers that (approximately) enjoy global optimality properties.
Given the success of RL and MPC individually, there is increasing interest in combining the two (e.g., \cite{mesbah2022fusion,reiter2025SynthesisModel,lawrence2025view}).
As a result, there are several possible ways in which RL and MPC can interact successfully.
This paper identifies two prominent approaches and evaluates them under the local-global framework; briefly, one approach takes a \emph{modular} view towards learning MPC, while the other adopts an \emph{all-in-one} perspective. 
To this end, we elucidate future research priorities towards automated and safe decision-making under uncertainty.

\section{Stochastic Optimal Control}

    \subsection{Markov Decision Process}

    A Markov decision process (MDP) serves as a useful paradigm for solving sequential decision-making problems.
    An MDP consists of an \emph{agent} interacting with an \emph{environment}.
    The environment comprises four components:
    \begin{enumerate}
        \item a state space with states $s \in \mathcal{S} \subseteq \mathbb{R}^{n}$;
        \item an action space with actions $a \in \mathcal{A} \subseteq \mathbb{R}^{m}$;
        \item a Markov kernel $p: \bar{\mathcal{S}} \times \mathcal{A} \times \mathcal{S} \rightarrow [0,1]$ describing the state transition dynamics;       
        \item and a reward function $r: \mathcal{S} \times \mathcal{A} \rightarrow \mathbb{R}$.
    \end{enumerate}
    The agent carries out the construction and deployment of a \emph{policy} $\pi: \mathcal{S} \times \mathcal{A} \rightarrow [0,1]$.
	Namely, at any state $s \in \mathcal{S}$, the agent produces an action $a \sim \pi(a | s)$.
	The action influences the environment, leading to a reward $r(s,a)$ and successive state $s' \sim p(s' | s, a)$. 
	This process is often indexed in time, leading to a sequence:
	\begin{equation}
		\{ s_0, a_0, r_0, s_1, \ldots s_t, a_t, r_t, s_{t+1}, \ldots \},
	\label{eq:sarsa}
	\end{equation}
	where $s_{0} \sim p(s_{0})$ and $r_t = r(s_t, a_t)$.

    The objective of the agent can be described as
    \begin{equation} \label{eqn: MDP objective}
        J(\pi) = \mathbf{E}_{\pi} \biggl\{ \sum_{t=0}^{\infty} \gamma^{t} r(s_{t},a_{t}) \biggr\},
    \end{equation}
    with discount factor $\gamma \in (0,1)$, to ensure convergence of the sum, serving as incentive for immediacy as opposed to delayed action.
    In \cref{eqn: MDP objective}, the expectation is taken with respect to trajectories in \cref{eq:sarsa}, where the actions are chosen according to $\pi$.
    Hence, solving the MDP involves obtaining an optimal policy $\pi^{\star}$ as
    \begin{equation} \label{eqn: optimal policy}
        \pi^{\star} = \argmax_{\pi \in \Pi} J(\pi).
    \end{equation}
    However, solving \cref{eqn: optimal policy} is challenging for multiple reasons \cite{bertsekas2012dynamic}. 
    First, searching the space of all feasible Markov kernel policies described by $\Pi$ is an intractable problem.
    Second, evaluating $J(\pi)$ exactly is complicated by the expectation $\mathbf{E}_{\pi}$.
    Third, the transition dynamics and reward function may be unknown.

    \subsection{Dynamic Programming}

    While solving \cref{eqn: optimal policy} may not always be tractable, dynamic programming exploits the MDP structure to guide the search for an optimal policy.
    To this end, it is convenient to introduce the state-action value function $Q^{\pi}: \mathcal{S} \times \mathcal{A} \rightarrow \mathbb{R}$ as
    \begin{equation} \label{eqn: Q-function}
        Q^{\pi}(s,a) = \mathbf{E}_{\pi} \biggl\{ \sum_{t=0}^{\infty} \gamma^{t} r(s_{t},a_{t}) \bigg| s_{0}=s, a_{0}=a \biggr\},
    \end{equation}
    for fixed $\pi \in \Pi$ and $\gamma \in (0,1)$.
    The state-action value function, or $Q$-function, corresponds to the expected future reward if the agent starts in state $s$, takes action $a$, and follows its policy $\pi$ from then on.
    The optimal $Q$-function $Q^{\star}$ describes the maximum expected future reward for any state-action tuple $(s,a) \in \mathcal{S} \times \mathcal{A}$, that is,
    \begin{equation} \label{eqn: optimal Q-function}
        Q^{\star}(s,a) = \max_{\pi \in \Pi} Q^{\pi}(s,a).
    \end{equation}
    The optimal $Q$-function is unique and obtained from the Bellman optimality equation \cite{bellman_adaptive_1959}
    \begin{equation} \label{eqn: Bellman optimality equation}
        Q^{\star}(s,a) = r(s,a) + \gamma\mathbf{E}_{s' \sim p(s'|a,s)}\biggl\{ \max_{a'} Q^{\star}(s',a')\biggr\}.
    \end{equation}

    The Bellman equation provides a recursive formula to solve for the optimal $Q$-function.
    In particular, \cref{eqn: Bellman optimality equation} is a fixed-point equation $Q^\star = \mathcal{T} Q^\star$, where $\mathcal{T}$ is a contraction mapping \cite{lewis_reinforcement_2012}, such that the optimal $Q$-function is a uniquely determined fixed-point via Banach fixed-point theorem \cite{banach_sur_1922}.
    Due to this fixed-point relationship, the optimal $Q$-function can be obtained by iterating 
    \begin{equation}
    	\mathcal{T} Q^{(k)} = Q^{(k+1)} \to Q^\star \quad \text{as } k \to \infty.
    	\label{eq:VI}
    \end{equation}
    Once a fixed-point is (nearly) achieved, an optimal policy $\pi^{\star}$ can be inferred by taking ``greedy'' actions with respect to the optimal $Q$-function
    \begin{equation} \label{eqn: greedy optimal policy}
        \pi^{\star}(s) = \argmax_{a} Q^{\star}(s,a).
    \end{equation}
    Accordingly, a policy $\pi^{\star} \in \Pi$ is optimal if and only if $Q^{\star}(s,a)$ satisfies the Bellman optimality equation \cref{eqn: Bellman optimality equation} for all $(s,a) \in \mathcal{S} \times \mathcal{A}$.

    The dynamic programming solution is computationally challenging for several reasons.
    For continuous state and action spaces, the $Q$-function comes from an infinite-dimensional function space and, as a consequence, the policy defined by \cref{eqn: greedy optimal policy} can become intractable.
    Furthermore, the Bellman equation requires evaluation of multidimensional integrals.
    Although \cref{eqn: Bellman optimality equation} can be approximated over a discretized space, the number of states grows exponentially in the state dimension for a fixed resolution.
    Thus, the so-called ``curse of dimensionality'' makes solving \cref{eqn: optimal policy} using dynamic programming intractable for many practical applications \cite{bellman_applied_1962}. 
    Nonetheless, \cref{eqn: Bellman optimality equation} is a useful theoretical paradigm for informing practical approximations.

\subsection{Reinforcement Learning}

RL refers to a class of algorithms for solving \cref{eqn: optimal policy} \cite{sutton_reinforcement_2018}.
Crucially, RL looks to learn optimal policies through interactions with its environment in a ``model-free'' manner; however, these methods may also be applied to a simulation environment to boost efficiency \cite{sutton1991dyna,recht2019tour,frauenknecht2024Trustmodel}.
While RL does not require an exact model of the environment, many algorithms draw from dynamic programming, leading to so-called \emph{value-based} approaches aimed at solving \cref{eqn: Bellman optimality equation} \cite{bertsekas_neuro-dynamic_1996}.
Other approaches target \cref{eqn: optimal policy} directly, leading to \emph{policy-based} algorithms \cite{grondman2012survey}.

\begin{itemize}
    \item {\bf Value-based.}\quad This approach takes advantage of the MDP structure by training a $Q$-function parameterization $Q_\phi$ to approximately satisfy the Bellman optimality equation \cref{eqn: Bellman optimality equation}. While there are many algorithmic intricacies underlying various value-based approaches, the basic objective is
        \begin{multline} \label{eqn: off-policy TD}
            \min_{\phi} \mathbf{E}_{(s,a,r,s')}\Bigg\{ \Big( Q_{\phi}(s,a) \\ - \left(r + \gamma \max_{a'} Q_{\phi}(s',a')\right)\Big)^2 \Bigg\},
        \end{multline}
        where the expectation is over sampled tuples $(s,a,r,s')$ from the environment. In this way, \cref{eqn: off-policy TD} is a loss based on the left- and right-hand sides of \cref{eqn: Bellman optimality equation}. A policy is then derived from $Q_{\phi}$ as in \cref{eqn: greedy optimal policy}. 
    \item {\bf Policy-based.}\quad Another strategy is to directly parameterize the policy as $\pi_\theta$. By considering trainable parameters $\theta$, such as weights of a neural network, \cref{eqn: optimal policy} can be recast over the parameter space as
        \begin{equation} \label{eqn: policy search}
            \max_{\theta} J(\pi_{\theta}).
        \end{equation}
        After all, the policy $\pi_\theta$ interacts with the environment, so directly parameterizing it is an appealing option. Typically, \cref{eqn: policy search} is solved through gradient-based schemes of the form
        \begin{equation} \label{eqn: gradient step}
        	\theta \leftarrow \theta + \alpha \nabla J(\pi_\theta),
        \end{equation}  
        where $\nabla J$ is generally estimated using policy gradient theorem \cite{sutton1999policy,silver2014deterministic}.
\end{itemize}

In practice, value-based and policy-based strategies are often combined, leading to \emph{actor-critic} RL methods \cite{konda1999actor,lillicrap2015continuous,haarnoja2018soft}. 
The actor refers to the policy and the critic refers to the value function.
Combining these two strategies allows for learning from a predefined policy class using one-step transitions.
This is an important property, as value-based approaches alone lead to policies of the form in \cref{eqn: greedy optimal policy}, which may be prohibitively expensive unless special care is taken.
Additionally, learning from one-step transitions is also an appealing benefit, as it enables online parameter updates, of the actor or critic, directly aimed at the optimal value function;
see \cite{grondman2012survey,haarnoja2018soft} for a detailed overview of the actor-critic framework and deep RL, respectively.


\section{The Local-Global Interface in Optimal Control}
\label{sec:local-global}

The Bellman optimality equation contains two insights:
\begin{itemize}
	\item A \emph{global} property that characterizes optimal trajectories over the whole state space.
	\item A \emph{local} strategy for interacting with the environment at a given state.
\end{itemize}
More concretely, \cref{eqn: Bellman optimality equation} can be understood entirely in terms of states since
\begin{multline} \label{eq:statevalue}
    \max_{a} Q^{\star}(s,a) \\ = \max_{a} \left\{ r(s,a) + \gamma\mathbf{E}_{s' \sim p(s'|a,s)}\biggl\{ \max_{a'} Q^{\star}(s',a')\biggr\}\right\}.
\end{multline}
\Cref{eq:statevalue} converts the original problem of determining an optimal policy, i.e., \eqref{eqn: optimal policy}, into a one-step optimality condition over state trajectories.
Consequently, \cref{eq:statevalue} must hold for all states $s \in \mathcal{S}$.
Hence, at a given $s \in \mathcal{S}$, the agent can construct an optimal trajectory online.
The left-hand side of \cref{eq:statevalue} instructs the agent simply to maximize $Q^{\star}$; the right-hand side assures the agent that, no matter how the environment uncertainty plays out, maximizing $Q^{\star}$ provides the best actions.

While our discussion assumed $Q^\star$ was known, the local-global interface is a general concept in stochastic optimal control, which also applies to the RL setting of iteratively approximating $Q^\star$.
In the following, we introduce a class of function approximators that is particularly well-aligned with the local-global concept.

\subsection{Optimization as Function Approximation}
\label{subsec:optFA}

To solve \cref{eqn: Bellman optimality equation}, the $Q$-function must be approximated.
By introducing a function approximator, the search space is reduced from an intractable function space to a set of parameters whose size can be controlled.
This new search space can then be designed with desirable attributes, such as differentiability.
Additionally, the requirement that \cref{eqn: Bellman optimality equation} holds for all $(s,a) \in \mathcal{S} \times \mathcal{A}$ is relaxed to hold in an approximate sense over a subset of $\mathcal{S} \times \mathcal{A}$.
While there are numerous choices of function approximators, including neural networks as a popular choice, we argue that parametric optimization is uniquely suited for solving \cref{eqn: Bellman optimality equation} and \cref{eqn: greedy optimal policy}.

In particular, we consider MPC as a specific class of parametric optimization.
An MPC agent considers the following value structure\footnote{To maintain consistency with the MDP maximization problem \cref{eqn: optimal policy}, we simply write $Q_\phi = - Q_\phi^\text{MPC}$.}
\begin{subequations} \label{eqn: Q-OCP}
    \begin{align}
        Q^{\text{MPC}}(s,a) = \min_{u_{0:H-1|t}} &~ \sum_{k=0}^{H-1} \ell(x_{k|t},u_{k|t}) + V(x_{H|t}), \label{eqn: Q-OCP objective} \\
                         \text{s.t.} &~ x_{k+1|t} = f(x_{k|t},u_{k|t}), \label{eqn: Q-OCP model} \\
                                     &~ g(x_{k|t},u_{k|t}) = 0, \label{eqn: Q-OCP equality constraint} \\
                                     &~ h(x_{k|t},u_{k|t}) \leq 0, \label{eqn: Q-OCP inequality constraint} \\
                                     &~ \forall k=0,\dots,H-1, \label{eqn: Q-OCP all time} \\
                                     &~ x_{0|t} = s, u_{0|t} = a, \label{eqn: Q-OCP initial state action}
    \end{align}
\end{subequations}
where $\ell$ is a stage cost, $V$ is a terminal cost, $f$ is a dynamic model, and $g$ and $h$ impose additional constraints. 
We write $Q_{\phi}^{\text{MPC}}$ to refer to a set of learnable parameters inside \cref{eqn: Q-OCP} related to the costs, model, or constraints.
For instance, we write $f_\phi$ if some components of $f$ are adjustable.

MPC operates in a receding-horizon fashion \cite{rawlings2017model}.
That is, at each sampling time, an action is designed with the optimization-based policy
\begin{equation}
    \pi^{\text{MPC}}_{\phi}(s) := \argmin_{a} Q^{\text{MPC}}_{\phi}(s,a) = u^{\star}_{0|t}.
    \label{eq:pi-ocp}
\end{equation} 
The optimal action $u^{\star}_{0|t}$ is then applied to the environment, and optimization \cref{eq:pi-ocp} is recomputed at the ensuing state.
This process of local replanning confers feedback and, therefore, some degree of robustness against uncertainties that  exist in the model and environment.\footnote{Uncertainties can be explicitly included in the design of MPC to provide robustness guarantees, leading to many variations of the nominal MPC formulation presented here for simplicity; see \cite{bemporad2007robust,mesbah2016stochastic}.} Note that the relationship between \cref{eqn: Q-OCP,eq:pi-ocp} is implicit, as it is not available in closed form. 

It follows that optimization as a function approximator provides a straightforward interface for integrating system knowledge into the policy structure.
Identifying $Q^{\text{MPC}}_{\phi}$ as \cref{eqn: Q-OCP}, one can embed prior models of the reward in objective \cref{eqn: Q-OCP objective}, dynamic models or other physical constraints in \cref{eqn: Q-OCP model,eqn: Q-OCP equality constraint}, and/or actuator limits and safe state regions in \cref{eqn: Q-OCP inequality constraint}, while being able to account for uncertainties within these models. Hence, the benefits of choosing the maps provided by parametric optimization for function approximation of $Q$ and $\pi$ can be attributed to their equation-oriented structure. 
These benefits include:
\begin{enumerate}
    \item \textbf{Modularity.}\quad \Crefrange{eqn: Q-OCP objective}{eqn: Q-OCP inequality constraint} allow for independent learning or design of individual components. Doing so can simplify learning with the inclusion of physics-based models, or other prior models, which can bring forth a reduction of learnable parameters in function approximators of $Q$ and $\pi$. Moreover, practical design requirements can be directly encoded through the cost functions and constraints.
    \item \textbf{Safety.}\quad Through inclusions of \cref{eqn: Q-OCP model,eqn: Q-OCP equality constraint,eqn: Q-OCP inequality constraint}, and possibly uncertainties, the parameterization readily accommodates (robust) constraint satisfaction. This aspect makes \cref{eqn: Q-OCP} well suited for function approximation in safety-critical applications, or when assurances such as stability or robustness are required.
    \item \textbf{Interpretability.}\quad Provided a comprehensible set of equations for \cref{eqn: Q-OCP objective,eqn: Q-OCP model,eqn: Q-OCP equality constraint,eqn: Q-OCP inequality constraint}, the clear specification of objective and constraints enhances understanding for how optimal decisions and values are obtained.           
\end{enumerate}
Thus, an effective prior for the MDP structure embedded in a parametric optimization can significantly simplify learning as opposed to selecting a highly-parameterized function approximator, such as a neural network, for $Q$ or $\pi$, which does not readily accommodate prior knowledge in its parametrization \cite{gros_data-driven_2020}.

\subsection{Local Actions and Global Optimality}

\begin{figure}[t!]
  \centering
  \includegraphics[width=0.85\linewidth]{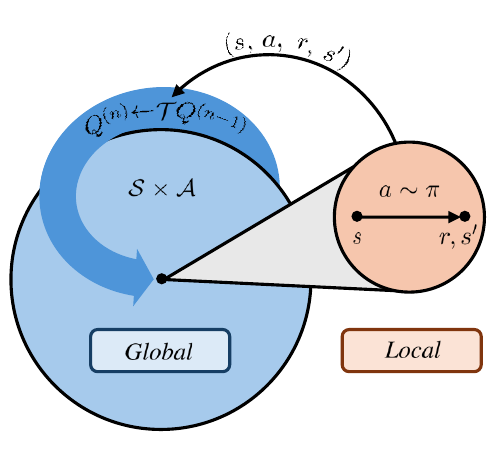}
  \caption{
  The local agent's interactions with the environment provide useful feedback towards improving the global MDP solution.
  The global $Q$-function should satisfy the Bellman optimality condition for all state-action pairs in $\mathcal{S} \times \mathcal{A}$.
  However, feedback at any given $(s,a)$ requires the local agent to visit these regions, using its policy $\pi$, as represented by the fragment of its trajectory $(s,a,r,s')$.
  Thus, improvements to the global solution are driven via local actions.
  }
  \label{fig:local-global_concept}
\end{figure}

\begin{figure}[t!]
  \centering
  \includegraphics[width=0.9\linewidth]{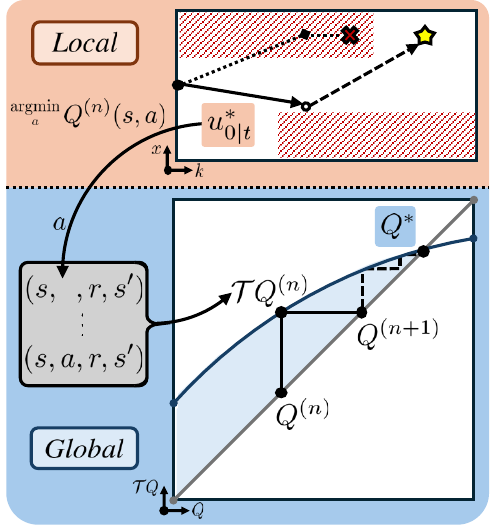}
  \caption{
  An MPC-based function approximator is well-aligned with the local-global interface. 
  Top: The local agent designs a coordinated sequence of actions to reach the target state, executing $u_{0|t}^{\star}$ as action $a$.
  As opposed to hastily taking direct action towards the target state, the agent moves away from the target to later satisfy the state constraints (dashed in red). 
  Bottom: In the global sense, a rich MPC parameterization can effectively approximate the optimal $Q$-function $Q^\star$.
  In theory, this approximation can be obtained by passing the global agent through the contraction mapping $\mathcal{T}$.
  However, $\mathcal{T}$ can only be approximately evaluated using observations $(s,a,r,s')$ obtained from the local agent's interactions.
  }
  \label{fig:local-global_detail}
\end{figure}

Although solving \cref{eqn: Bellman optimality equation} is generally intractable, approximating the \textit{global} MDP solution using a $Q$-function parameterization can provide a practical pathway to obtaining near-optimal policies \cite{sutton_reinforcement_2018}. 
This global approximator can aid in the design of optimal actions, similar to \cref{eqn: greedy optimal policy}, defining a policy useful for \textit{local} online decision-making.
This idea is illustrated in \Cref{fig:local-global_concept}, with additional detail pertaining to MPC-based approximators in \Cref{fig:local-global_detail}.
We further delineate the global and local aspects in the context of MPC-based approximators below.

\textbf{Global}.\quad
For a chosen $Q$-function approximator $Q_{\phi}$ to be viable, there should exist a set of parameters $\phi^{\star}$ whose associated $Q$-function approximately match that of the global optimum, that is, $Q_{\phi^{\star}}(s,a) \approx Q^{\star}(s,a)$.
The global (approximate) solution $Q_{\phi^{\star}}(s,a)$ is verifiable based upon its fulfillment of the Bellman optimality condition in \cref{eqn: Bellman optimality equation} for all $(s,a) \in \mathcal{S} \times \mathcal{A}$.
Because of this, $Q_{\phi}$ must model the entire domain, representing some ``generalist'' sense of value.

Defining $Q_{\phi}$ in terms of \cref{eqn: Q-OCP} is intuitively appealing in light of the benefits of optimization-based function approximators---modularity, reliability, interpretability---as outlined above.
Namely, the task of designing a global, generalizable $Q$-function can be initialized with independent building blocks.
These may include state-action constraints, specifying what states and actions the agent must avoid; smooth cost functions, shaping the agent's path to a goal; and physical models, limiting the agent to plausible scenarios.
These pieces can all be grouped into a set of parameters $\phi$.

While $Q_{\phi}$ contains modular components and can provide a good initialization, improving these components is still nontrivial. 
Constraints may be imperfect or softened, the cost may under- or over-penalize decision variables, and so on.
These issues can be corrected by adapting $\phi$ through interactions with the system.
\cref{eq:VI} would suggest this is simply a matter of churning through $Q$-functions over $\mathcal{S} \times \mathcal{A}$.
The resulting $Q_\phi$ should certainly be calibrated with broad coverage in $\mathcal{S} \times \mathcal{A}$.
In reality, however, a localized perspective is also required.

\textbf{Local}.\quad 
While the purpose of $Q_{\phi}$ is to model the entire global domain, generating actions is a local process for each state encountered online.
For these local decisions to optimize the agent's objective in \cref{eqn: MDP objective}, they should maximize the global $Q$-function as in \cref{eqn: greedy optimal policy}.
As such, given the $Q$-function approximator in \cref{eqn: Q-OCP}, an approximate version of \cref{eqn: greedy optimal policy} can be solved to construct a ``local agent'' defined by the online refinement of $Q_{\phi}$ at each newly encountered state. That is, the policy is identified with \cref{eq:pi-ocp}.


In dynamic programming, this can be thought of as a step of policy improvement and, as a consequence of the policy improvement theorem, $\pi_{\phi}$ is guaranteed to be as good as, or better than, the policy best represented by the learned $Q_{\phi}$ \cite{sutton_reinforcement_2018}.
Moreover, the deployment of $\pi^{\text{MPC}}_\phi$ in \cref{eq:pi-ocp} entails online optimization, thereby accounting for constraints in its design.
Therefore, the local agent aims to operate in a ``desirable'' region of $\mathcal{S} \times \mathcal{A}$.
This focused behavior can provide feedback in regions where the global agent may have made errant value estimates.
Through its interactions with the environment, the local agent iteratively drives refinements over the global agent by generating trajectories useful towards recalibrating the global agent.

\section{Practical Complications in Local-Global Learning}
\label{sec:compete}

\Cref{sec:local-global} gives a somewhat idealized overview of the local-global interface in optimal control.
In this section, we highlight some of the main practical challenges that arise in this framework, as they pertain to learning optimization-based function approximators.

\subsection{Local-Global Optimization Requirements}


The optimization in \cref{eqn: greedy optimal policy} is generally intractable, potentially undermining the real-time deployment of a local, optimization-based policy.
For the local-global interface to be realizable for online optimization, one should be mindful that $Q_{\phi}$ be
\begin{enumerate}
    \item rich enough to approximate the \textit{global} MDP solution, i.e., satisfy \cref{eqn: Bellman optimality equation}; 
    \item conducive to \textit{local} online refinement of the approximation, i.e., solving \cref{eqn: greedy optimal policy}.
\end{enumerate}
Effectively fulfilling both can be accomplished by incorporating prior knowledge of the MDP structure into the parameterization of $Q^{\text{MPC}}_{\phi}$ in \cref{eqn: Q-OCP}, and subsequently exploiting said structure in optimizing $\pi^{\text{MPC}}_\phi$ in \cref{eq:pi-ocp}.

It has been shown in \cite{gros_data-driven_2020} that the optimal $Q$-function can be recovered exactly by an MPC-based function approximator provided that the parameterization is rich enough.
While a parameterization satisfying the global requirement may exist, how to acquire said parameterization is an open question.
As a consequence, one cannot verify if the chosen MPC is overly parameterized or simplistic such that $Q^{\star}$ is excluded.
The former hinders local decision-making in real-time, while the latter precludes global optimality.
However, \cite{gros_data-driven_2020} and \cite{gros_learning_2022} show that global optimality is achievable when incorporating prior knowledge of the MDP (even if imperfect) and learning a rich cost function.

Furthermore, the local requirement of practical online optimization can be addressed using efficient nonlinear optimization solvers such as IPOPT \cite{wachter2006implementation}.
Advances in numerical optimization have created significant opportunities to construct the implicit map of \cref{eq:pi-ocp} in real-time as new states are encountered.
In this way, using large constrained nonlinear optimization problems as rich $Q$-function parameterizations does not significantly inhibit the online access to local agent's decision-making policy or value function approximation.
Finally, locally refining $Q^{\text{MPC}}_{\phi}$ towards the redesign of actions serves to improve the policy and provides a degree of robustness to errors in approximation.

\subsection{Exploration versus Exploitation}

The true transition dynamics $p$, and sometimes reward $r$, in \cref{eqn: Bellman optimality equation} are generally unknown.
This makes it infeasible to estimate $Q^\star$ over the whole state space as \cref{eq:VI} would suggest.
Therefore, acquiring and verifying $Q_{\phi^{\star}}$ relies on limited observations of the form $(s,a,r,s')$ that stem from the agent's interactions with the environment.
These samples in combination with the known condition for the Bellman equation can structure the search for $\phi^{\star}$: modifying $Q_{\phi}$ to best satisfy \cref{eqn: Bellman optimality equation} in the observed regions of the state space.

This notion of using samples to calibrate $Q_{\phi}$ towards the global Bellman equation leads to the question of how the local agent should go about acquiring information from the environment.
If it only operates in a greedy fashion, as in \cref{eq:pi-ocp}, the agent runs the risk of optimizing \cref{eqn: off-policy TD} in a narrow region of the state space.
Alternatively, if the agent deviates from its greedy policy and tries novel actions, it may encounter more fruitful state-action sequences, which can then be used to fine-tune the Bellman estimate in \cref{eqn: off-policy TD}.
This tension is known as the \emph{exploration-exploitation tradeoff} \cite{sutton_reinforcement_2018}.


Although there are many exploration strategies in the RL literature \cite{powell2011ApproximateDynamic}, they are somewhat ad hoc. 
Indeed, a natural question is to what extent exploration is ``optimal'' for a given task.
Dual control \cite{feldbaum1960dual} gives a formal answer to this question through the dynamic programming framework, wherein uncertainty in the environment is mitigated to the degree required by the control objective \cite{ wittenmark1995adaptive}.
Unsurprisingly, dual control is generally intractable since it targets the global Bellman equation in \cref{eqn: Bellman optimality equation}.
However, it has inspired many approximate dynamic programming and practical MPC schemes that actively probe the environment, thereby adapting to environmental changes and mitigating uncertainty; see \cite{mesbah2018stochastic} for an overview.
In the context of RL, the MPC structure can be used to induce coordinated sequences of exploration actions.
This can be achieved by modifying the cost.
For instance, \cite{zanon2020safe} added a random penalty term to the cost, which perturbs the decision variables while maintaining feasibility.
\cite{lowrey2018plan} incorporated an ensemble of terminal value functions into the objective in such a way to encourage the MPC agent to explore regions in the state space with large disagreement among the ensemble.

\section{Case Studies on Interfacing MPC and RL}

We present two case studies, each designed to illustrate different aspects of the local-global interface in optimal control.
The first one demonstrates a modular approach to designing MPC-based local agents, wherein an RL-learned value function is used as a terminal cost.
A full explanation of this example is provided in \cite{lawrence2025view}. Moreover, the general idea behind this example is demonstrated in a number of works; see, for example, \cite{bertsekas2022lessons, lee2001NeurodynamicProgramming, lee2009ApproximateDynamic, zhong2013Valuefunction, lowrey2018plan, farshidian2019DeepValue}, and \cite{bhardwaj2020BlendingMPC}. The second example demonstrates an all-in-one strategy in which RL directly updates the MPC parameters; further details can be found in \cite{banker_gradient-based_2025}.
Again, this example is representative of a dominant approach for combining RL and MPC; see, for example, \cite{gros_data-driven_2020, gros_learning_2022, chen_gnu-rl_2019, amos_differentiable_2019, romero2024ActorCriticModel, hansenTemporalDifferenceLearning2022}, and \cite{anand_painless_2023}.

\subsection{Value Function-Augmented MPC}
\label{subsec:vf}

A benefit of the MPC structure in \cref{eqn: Q-OCP} is the ability to design its individual components independently.
For example, the dynamic model in \cref{eqn: Q-OCP model} may come from prior physical knowledge or system identification.
Such models are beneficial for theoretical aspects such as stability \cite{mayne2000Constrainedmodel} and efficient learning under certainty equivalence control \cite{maniaCertaintyEquivalenceEfficient2019}. 
However, they are not necessarily optimal for control, as system identification and dynamic programming have different objectives \cite{gevers2005identification}.
On the other hand, the MPC structure contains other terms that can be modified in a performance-driven fashion.
Namely, the terminal cost $V$ in \cref{eqn: Q-OCP objective} is a suitable object to learn via RL.

The terminal value function in MPC approximates the optimal cost-to-go at the final predicted state.
In nominal MPC, it is often designed based on the unconstrained LQR solution \cite{lee2011Modelpredictive}.
Here, this idea is taken as inspiration for an RL training scheme.
Specifically, an RL agent learns $V$ through offline training in a complex simulation environment that accounts for system uncertainties.
In doing so, the RL agent looks to find a global value function most immediately concerned with rapidly maximizing reward in \cref{eqn: MDP objective}.
Similar to the LQR setting, the RL agent is not concerned with constraints in this offline setup. Instead, it attempts to achieve the control objective at a rate largely determined by the discounting $\gamma$.

Of course, in many practical applications, there exist limitations or constraints within the state-action space, and these may be in conflict with the goal of rapidly maximizing \cref{eqn: MDP objective}.
If global learning does not encourage respecting system constraints, this can result in the agent taking severe, possibly unsafe actions, traversing dangerous regions of the state space, or exhibiting otherwise undesirable behavior for the sake of quickly achieving the goal.
With this in mind, the RL agent is not directly deployed for online control.
Instead, the RL-learned value function is incorporated into an MPC agent.
This value function-augmented MPC agent accounts for constraints in its design, while benefiting from the flexibility of RL training. Such flexibility manifests in a couple of ways.
First, the offline RL agent can learn under complex objectives, such as binary rewards, that may be impractical for an MPC agent.
Second, the RL simulation environment may be complex and uncertain. In such cases, long-term MPC predictions may become computationally intractable, while the RL-learned value function summarizes all the planning within $V$.

\begin{figure}[t!]
	\includegraphics[width=\linewidth]{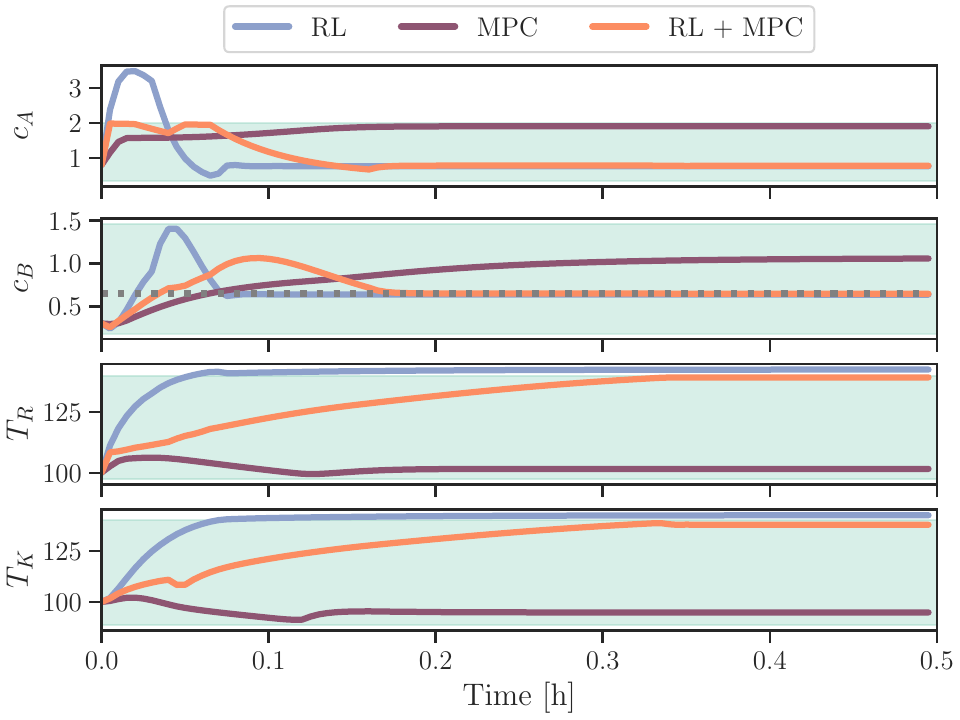}
	\caption{Closed-loop state profiles of a continuous stirred tank reactor for three optimal decision-making agents: RL, MPC, and RL+MPC \cite{lawrence2025view}. The MPC agent satisfies the constraints (shaded area), but never achieving the control target (dashed $c_B$ value). The RL agent immediately violates the constraints, but ultimately reaches the control target. The RL+MPC (value function-augmented MPC) agent tames the RL agent's trajectory, satisfying the constraints, while also eventually reaching the target. The state variables are concentrations $c_A$ and $c_B$, as well as reactor and coolant temperatures $T_R$ and $T_K$.}
	\label{fig:cstr_rlmpc_profile}
\end{figure}

This strategy is demonstrated on a continuous stirred tank reactor environment, a common benchmark in process control \cite{klatt1998Gainschedulingtrajectory,fiedler2023DompcFAIR,bloor2024PCGymBenchmark}; see \cite{lawrence2025view} for full experimental details.
In \Cref{fig:cstr_rlmpc_profile}, the RL agent has no knowledge of the desired operating region, leaving the constraints in favor of quickly achieving its goal.
Meanwhile, the MPC agent stays inside the operating region, but at the cost of never tracking the desired concentration.
This MPC agent is taken from \cite{fiedler2023DompcFAIR}; it indeed performs very well in a narrow region of the state space, but can run into the issue shown here.
Finally, the value function-augmented MPC agent (RL+MPC) is able to effectively balance the benefits of both RL and MPC, achieving its goal while remaining within the safe operating region. This example illustrates how a global RL agent can be trained offline in a goal-driven fashion and then refined locally via MPC.
Rather than training the RL agent with penalties in its reward function, which requires careful fine-tuning, constraints are enforced directly through MPC to more effectively account for the (possibly evolving) requirements of online control.



%

\subsection{MPC as a Function Approximator}
\label{subsec:fa}


An alternative strategy for uniting RL and MPC is to treat the implicit maps $Q_{\phi}$ and/or $\pi_{\phi}$ in their entirety as function approximators whose individual components are freely adapted to maximize a singular RL objective.
Due to the local and global perspectives offered by an MPC-based approximator, any component of the MPC agent---the parametric models defining the objective \cref{eqn: Q-OCP objective} and constraints \cref{eqn: Q-OCP model,eqn: Q-OCP equality constraint,eqn: Q-OCP inequality constraint}---is modifiable with value-based or policy-based approaches.
Following this strategy, $Q_{\phi}^\text{MPC}$ and $\pi_{\phi}^\text{MPC}$ are suitable for online learning in the sense that the agent learns from repeated interactions with the true system.
With repeated interactions, the agent can explore the true system to optimize its policy.
This is in contrast to learning-based control strategies that learn from a simulator, or a fixed historical dataset of the true system \cite{tan_sim--real_2018,witman2019sim,levine_offline_2020}.
Thus, online learning can avoid issues of distribution shift, limited coverage, and otherwise incomplete representation of the environment that come from optimizing to a fixed dataset \cite{levine_offline_2020}.
By interacting with the true system, it can be ensured that the agent’s learning is free from model bias, thereby guiding the MPC agent towards optimal behavior.

In this online, all-in-one training setup, all of the learnable parameters $\phi$ are updated in tandem using gradient-based update rules, typically in an unconstrained fashion as in \cref{eqn: gradient step}.
The gradient-based updates require calculating the sensitivities $\nabla_{\phi}\pi_{\phi}^\text{MPC}$ or $\nabla_{\phi}Q_{\phi}^\text{MPC}(s,a)$ for policy-based and value-based approaches, respectively.
Attaining the former requires differentiating the optimal solution of the parametric optimization \cite{dontchev_implicit_2014}, while the latter requires differentiating the Lagrangian \cite{rockafellar_directional_1984}.
The idea being that synchronous, successive updates to the MPC agent's components will steer the agent towards optimality.
This is appealing in that a single update rule applied to any combination of MPC components can yield a tailored, performance-oriented MPC agent.
Following this point, MPC as a function approximator is a straightforward strategy that can be effectively integrated with classical RL algorithms such as Monte Carlo methods and Q-learning \cite{chen_gnu-rl_2019,gros_data-driven_2020}.
These algorithms generally have simple update rules that can be readily implemented for MPC agents.
Furthermore, they are applicable to MPC agents spanning from simple convex formulations to nonlinear problems.

\begin{figure}[t!]
	\includegraphics[width=\linewidth]{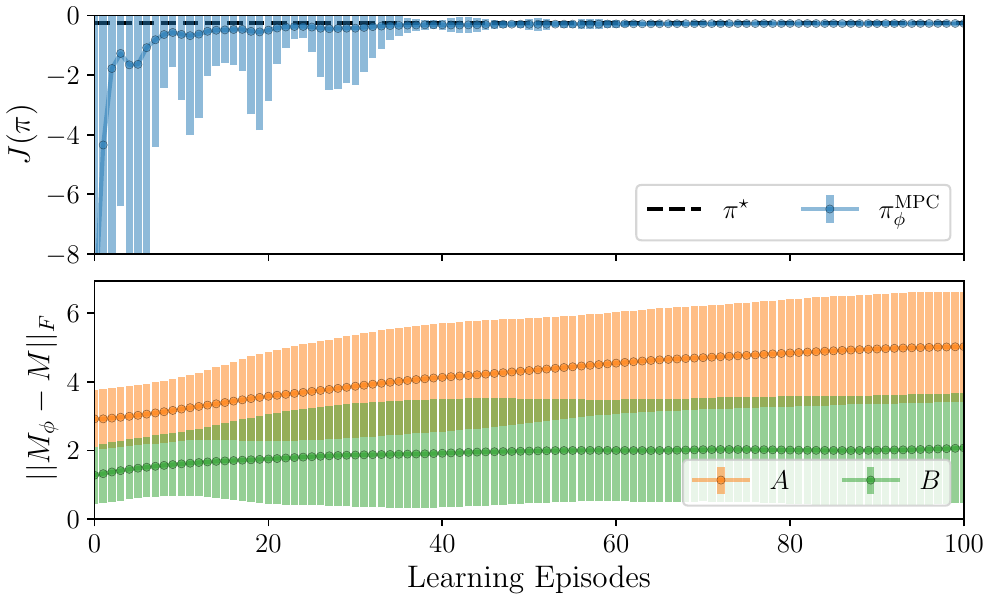}
	\caption{
    Episodic adaptation of the model of an MPC agent used as a function approximator \cite{banker_gradient-based_2025}. Top: Episodic return of MPC as a policy function approximator with learnable state-space model matrices $A$ and $B$. Initially, the observations of objective $J(\theta)$ exhibit large variance. After several learning episodes, the agent approaches the optimal $J(\theta^\star)$.
    Bottom: Mismatch between the MPC agent's state-space model matrices and those of the true system (denoted by $M_\phi$ and $M$, respectively) quantified in terms of the Frobenius norm $||\cdot||_{F}$. Notably, as the agent's performance improves, the model mismatch increases.
    Lines represent the mean of 100 runs with $\pm 2$ sample standard deviations shaded.
    }
	\label{fig:LQ_J-mismatch.pdf}
\end{figure}

To demonstrate these points, consider MPC as a function approximator for an agent's policy $\pi_{\phi}$ in an environment with linear dynamics and quadratic reward function.
The linear state-space model of the MPC agent is episodically updated in a gradient-based fashion, as in \cref{eqn: gradient step}, using the REINFORCE algorithm; details can be found in \cite{banker_gradient-based_2025}.
As shown in \Cref{fig:LQ_J-mismatch.pdf}, this online learning strategy yields control policies $\pi^{\text{MPC}}_{\phi}$  that gradually approach the optimal policy $\pi^{\star}$ through repeated interactions with the environment over several episodes. 
Notably, in the process of achieving this performance, the dynamic model learned by the MPC agent exhibits a gradual increase in mismatch for both matrices $A$ and $B$ (\Cref{fig:LQ_J-mismatch.pdf} bottom).
Rather than converging to the most accurate model for prediction, corresponding to minimal mismatch, the MPC agent learns a \emph{performance-oriented} model to achieve near optimal performance.
The all-in-one aspect of MPC as a function approximator in this example can be viewed as including system identification as part of \cref{eqn: policy search}.
This embodies the notion of identification for control \cite{gevers2005identification}, wherein the global objective dictates \emph{control-oriented} adaptation of the MPC policy parameters $\phi$.
That is, the local agent designs actions to adapt  $\phi$ in a performance-oriented manner. 


\subsection{Overall Evaluation}


Having exemplified the modular and all-in-one perspectives of the local-global interface in optimal control, we now evaluate their relative advantages and disadvantages.
We begin by discussing how each strategy treats the value structure $Q_{\phi}^\text{MPC}$ and policy $\pi_{\phi}^\text{MPC}$, as introduced in \cref{subsec:optFA}.
To this end, we outline the implications of learning these optimization-based approximators.
After which, in light of the complications laid out in \cref{sec:compete}, we describe how each strategy is suited for the local-global optimization requirements, broader integration with RL, and providing guarantees in exploration and exploitation.

\textbf{Treatment of MPC}.\quad
MPC as a function approximator disregards the individual functionality of the MPC agent's components in favor of treating $\pi^{\text{MPC}}_{\phi}$ and $Q^{\text{MPC}}_{\phi}$ as generic implicit functions.
As an example, the model in \cref{eqn: Q-OCP model} has the function of predicting the system's behavior, but its learnable parameters $\phi$ are manipulated without direct consideration of this purpose.
This generic treatment benefits from simplicity; a single update rule applies to all components, which effectively reduces algorithmic overhead.
Thus, the MPC agent is bestowed great freedom in the manipulation of $\pi^{\text{MPC}}_{\phi}$ and $Q^{\text{MPC}}_{\phi}$.
Although the learned implicit maps can approach optimal behavior, the associated components are not necessarily interpretable in isolation when following this strategy.
By ignoring their functional aspects, each component's contribution to the MPC agent's behavior is obfuscated.
An unintended consequence of this is that the RL updates to $\phi$ can affect the agent's behavior in a manner that is challenging to interpret.

In contrast, a value function-augmented strategy respects the functionality of each component and, thus, effectively leverages the modularity of an MPC agent.
By using RL and system identification to learn the value function and model, respectively, separation is established, which clarifies the role of each component within the overall MPC structure.
With modular updates, changes to the MPC agent's behavior are more predictable compared to the more generic and all-in-one treatment of $Q^{\text{MPC}}_{\phi}$ as an implicit function.
However, a value function-augmented strategy is not immediately equipped for learning all components of the MPC agent.
Therefore, parametric stage costs $\ell$ and constraints $g, h$ are fixed according to prior knowledge and are potentially tedious to design.
With this in mind, goal-conditioned RL \cite{andrychowicz2017HindsightExperience,eysenbach2022contrastive} aims to alleviate the burden of designing a reward function, which is also a promising avenue for designing MPC agents, as shown in \cite{lawrence2025view}.

\textbf{Optimization and Algorithmic Considerations}.\quad
MPC as a function approximator has been shown to be an effective strategy in ``simple'' environments (e.g., deterministic settings with small state-action spaces and immediate feedback), where certainty equivalence, linear dynamics, quadratic objectives, and other modest MPC formulations are typically successful.
While not richly parameterized, these modest formulations generally satisfy the local requirement of efficient online optimization, and they can be readily adapted by classical RL algorithms.
However, this strategy is less explored for environments with large state-action spaces, stochastic and uncertain settings, or highly nonlinear dynamics.
In these cases, simple MPC formulations may lack sufficient richness to effectively approximate the global solution, while classical RL algorithms, which have been shown to suffer from divergence, over-estimation, and high variance \cite{sutton_reinforcement_2018}, are ill-equipped.
Though there has been progress towards integrating machine learning models with MPC tools \cite{salzmann2024LearningCasADi}, which are increasingly versatile and user-friendly \cite{fiedler2023DompcFAIR}, the integration of MPC architectures within established RL libraries is lacking.

In the recent, yet dense, history of RL, novel algorithms addressing the latter issues have been developed \cite{schulman_high-dimensional_2018,kumar_stabilizing_2019}, but practically interfacing sufficiently rich MPC agents with these algorithms remains challenging.
Critically, existing RL ecosystems do not readily accommodate the batching and sensitivity propagation of nonlinear, robust/stochastic, or otherwise elaborate MPC agents.
As a consequence, interfacing nontrivial MPC agents with RL can become an expensive and inefficient process in engineering and computation.

Nonetheless, by once again leveraging MPC agent's modularity, the value function-augmented strategy effectively sidesteps the issue of integrating the MPC agent with the RL ecosystem.
Within the RL ecosystem, the RL agent is flexible to the choice of algorithm and, after learning in simulation, only the RL-learned value function is ported to the MPC agent.
Thus, there is no need for batching the MPC agent, or obtaining its sensitivities.
However, one should note that the resulting RL-learned value function is suitable for local online refinement.
For example, a neural network parameterization can effectively approximate the global MDP solution, but it can impede the local design of actions due to being non-convex and beset with many local minima.
Additionally, depending on the duration of control intervals, real-time optimization of the network can be computationally prohibitive.
Yet, it is encouraging to see strides towards improving the integration of machine learning models (for value function representation) with frameworks for nonlinear optimization such as CasADi \cite{andersson2019casadi}; e.g., see \cite{salzmann2024LearningCasADi}.

\textbf{Guarantees in Exploration and Exploitation}.\quad
A consequence of unconstrained learning of MPC as a function approximator is that intermediate policies do not necessarily provide guarantees on safety or stability. 
Being all-in-one and performance-oriented, the learned model in \cref{eqn: Q-OCP model} may not be accurate in prediction if manipulated for the RL objective.
Accordingly, an inaccurate model complicates enforcing constraints on the MPC agent's trajectory, either in exploration or exploitation to reach its goal.
Because the agent learns from online interaction with the true system, this presents risks wherein the MPC agent can violate safety-critical state constraints or deploy an unstable policy.
\cite{gros_learning_2022} addressed this challenge with safe RL, a constrained optimization approach, but for a robust linear MPC setting.
Although a rich parameterization can recover optimal behavior with an inaccurate model, it is recognized that modest MPC formulations cannot be expected to surely fulfill this requirement.

Guarantees in exploration is less of a concern for value function-augmented strategies, where violating constraints in a simulation environment is of little consequence to safety.
With this in mind, value function-augmented strategies can allot the agent nearly unlimited ability to explore in the simulation environment.
However, high-fidelity simulators can be expensive to develop and are not available for every system. 
Furthermore, effective simulators almost always have missing or incomplete information of the real world.
Because of this reality gap, global optimality in simulation does not necessarily translate to optimal local actions on the true system \cite{tan_sim--real_2018}.
Rather, as the MPC agent exploits its simulation-based knowledge via online refinement of the value function, it may be misguided by mismatch between simulation and the true system.
On the other hand, recent model-based RL methods have been shown to match the performance of model-free approaches, thereby increasing sample efficiency and mitigating the sim2real gap \cite{janner2019WhenTrust,frauenknecht2024Trustmodel}; integration with MPC is then a promising option.
Additionally, incorporating uncertainty into the simulator can help address the sim2real gap \cite{lawrence2025view}, whereas online learning via the all-in-one approach avoids the challenge of transferring simulation-based knowledge to the true system \cite{banker_gradient-based_2025}.

\section{Conclusions}

The local-global perspective in optimal control holds significant promise towards developing interpretable control policies.
In particular, a local view is amenable to deriving optimization-based policies with a modular and interpretable structure, while a global view aims to steer these policies towards optimality under the Bellman optimality condition.
Due to the structure of optimization-based control policies, there are multiple ways in which the local-global framework can manifest itself, containing different advantages and disadvantages.
By evaluating two prominent approaches, we have outlined fruitful paths forward in pursuit of more intelligent and safe decision-making agents.





\bibliographystyle{IEEEtran}
\bibliography{refs}

\begin{thebibliography}{10}
\providecommand{\url}[1]{#1}
\csname url@rmstyle\endcsname
\providecommand{\newblock}{\relax}
\providecommand{\bibinfo}[2]{#2}
\providecommand\BIBentrySTDinterwordspacing{\spaceskip=0pt\relax}
\providecommand\BIBentryALTinterwordstretchfactor{4}
\providecommand\BIBentryALTinterwordspacing{\spaceskip=\fontdimen2\font plus
\BIBentryALTinterwordstretchfactor\fontdimen3\font minus \fontdimen4\font\relax}
\providecommand\BIBforeignlanguage[2]{{%
\expandafter\ifx\csname l@#1\endcsname\relax
\typeout{** WARNING: IEEEtran.bst: No hyphenation pattern has been}%
\typeout{** loaded for the language `#1'. Using the pattern for}%
\typeout{** the default language instead.}%
\else
\language=\csname l@#1\endcsname
\fi
#2}}

\bibitem{bellman_dp_1957}
R.~E. Bellman, \emph{Dynamic Programming}.\hskip 1em plus 0.5em minus 0.4em\relax Princeton, NJ: Princeton University Press, 1957.

\bibitem{sutton_reinforcement_2018}
R.~S. Sutton and A.~G. Barto, \emph{Reinforcement {Learning}: {An} {Introduction}}.\hskip 1em plus 0.5em minus 0.4em\relax Cambridge, MA: MIT Press, 2018.

\bibitem{powell2011ApproximateDynamic}
W.~B. Powell, \emph{Approximate Dynamic Programming: Solving the Curses of Dimensionality}, 2nd~ed., ser. Wiley Series in Probability and Statistics.\hskip 1em plus 0.5em minus 0.4em\relax Hoboken, NJ: Wiley, 2011.

\bibitem{bertsekas_neuro-dynamic_1996}
D.~Bertsekas and J.~Tsitsiklis, \emph{Neuro-{Dynamic} {Programming}}.\hskip 1em plus 0.5em minus 0.4em\relax Nashua, NH: Athena Scientific, 1996.

\bibitem{kalman1963theory}
R.~E. K{\'a}lm{\'a}n, ``The theory of optimal control and the calculus of variations,'' \emph{Mathematical Optimization Techniques}, vol. 309, p. 329, 1963.

\bibitem{bertsekas2012dynamic}
D.~Bertsekas, \emph{Dynamic Programming and Optimal Control: {{Volume I}}}, 3rd~ed.\hskip 1em plus 0.5em minus 0.4em\relax Nashua, NH: Athena Scientific, 1995.

\bibitem{bertsekas2022lessons}
------, \emph{Lessons from {{AlphaZero}} for Optimal, Model Predictive, and Adaptive Control}.\hskip 1em plus 0.5em minus 0.4em\relax Nashua, NH: Athena Scientific, 2022.

\bibitem{rawlings2017model}
J.~B. Rawlings, D.~Q. Mayne, and M.~Diehl, \emph{{Model Predictive Control: Theory, Computation, and Design}}.\hskip 1em plus 0.5em minus 0.4em\relax Santa Barbara, CA: Nob Hill Publishing, 2017.

\bibitem{schulman2017ProximalPolicy}
J.~Schulman, F.~Wolski, P.~Dhariwal, A.~Radford, and O.~Klimov, ``Proximal policy optimization algorithms,'' 2017, arXiv:1707.06347.

\bibitem{lee2011Modelpredictive}
J.~H. Lee, ``Model predictive control: {{Review}} of the three decades of development,'' \emph{International Journal of Control, Automation and Systems}, vol.~9, no.~3, pp. 415--424, 2011.

\bibitem{mesbah2022fusion}
A.~Mesbah \emph{et~al.}, ``Fusion of machine learning and {MPC} under uncertainty: What advances are on the horizon?'' in \emph{Proceedings of the American Control Conference}, Atlanta, 2022, pp. 342--357.

\bibitem{reiter2025SynthesisModel}
R.~Reiter \emph{et~al.}, ``Synthesis of model predictive control and reinforcement learning: {{Survey}} and classification,'' 2025, arXiv.2502.02133.

\bibitem{lawrence2025view}
N.~P. Lawrence, P.~D. Loewen, M.~G. Forbes, R.~B. Gopaluni, and A.~Mesbah, ``A view on learning robust goal-conditioned value functions: Interplay between {RL} and {MPC},'' 2025, arXiv:2502.06996.

\bibitem{bellman_adaptive_1959}
R.~Bellman and R.~Kalaba, ``On adaptive control processes,'' \emph{IRE Transactions on Automatic Control}, vol.~4, no.~2, pp. 1--9, 1959.

\bibitem{lewis_reinforcement_2012}
F.~L. Lewis, D.~Vrabie, and K.~G. Vamvoudakis, ``Reinforcement learning and feedback control: Using natural decision methods to design optimal adaptive controllers,'' \emph{IEEE Control Systems Magazine}, vol.~32, pp. 76--105, 2012.

\bibitem{banach_sur_1922}
S.~Banach, ``Sur les opérations dans les ensembles abstraits et leur application aux équations intégrales,'' \emph{Fundamenta Mathematicae}, vol.~3.

\bibitem{bellman_applied_1962}
R.~E. Bellman and S.~E. Dreyfus, \emph{Applied {Dynamic} {Programming}}.\hskip 1em plus 0.5em minus 0.4em\relax Princeton, NJ: Princeton University Press, 1962.

\bibitem{sutton1991dyna}
R.~S. Sutton, ``Dyna, an integrated architecture for learning, planning, and reacting,'' \emph{ACM Sigart Bulletin}, vol.~2, no.~4, pp. 160--163, 1991.

\bibitem{recht2019tour}
B.~Recht, ``A tour of reinforcement learning: The view from continuous control,'' \emph{Annual Review of Control, Robotics, and Autonomous Systems}, vol.~2, no.~1, pp. 253--279, 2019.

\bibitem{frauenknecht2024Trustmodel}
B.~Frauenknecht, A.~Eisele, D.~Subhasish, F.~Solowjow, and S.~Trimpe, ``Trust the model where it trusts itself - model-based actor-critic with uncertainty-aware rollout adaption,'' in \emph{Proceedings of the 41st International Conference on Machine Learning}, ser. Proceedings of Machine Learning Research, vol. 235, Vienna, 2024, pp. 13\,973--14\,005.

\bibitem{grondman2012survey}
I.~Grondman, L.~Busoniu, G.~A. Lopes, and R.~Babuska, ``A survey of actor-critic reinforcement learning: Standard and natural policy gradients,'' \emph{IEEE Transactions on Systems, Man, and Cybernetics, part C (applications and reviews)}, vol.~42, no.~6, pp. 1291--1307, 2012.

\bibitem{sutton1999policy}
R.~S. Sutton, D.~McAllester, S.~Singh, and Y.~Mansour, ``Policy gradient methods for reinforcement learning with function approximation,'' in \emph{Advances in Neural Information Processing Systems}, vol.~12.\hskip 1em plus 0.5em minus 0.4em\relax Denver: MIT Press, 1999.

\bibitem{silver2014deterministic}
D.~Silver, G.~Lever, N.~Heess, T.~Degris, D.~Wierstra, and M.~Riedmiller, ``Deterministic policy gradient algorithms,'' in \emph{Proceedings of the 31st International Conference on Machine Learning}, ser. Proceedings of Machine Learning Research, vol.~32, Beijing, 2014, pp. 387--395.

\bibitem{konda1999actor}
V.~Konda and J.~Tsitsiklis, ``Actor-critic algorithms,'' in \emph{Advances in Neural Information Processing Systems}, vol.~12.\hskip 1em plus 0.5em minus 0.4em\relax Denver: MIT Press, 1999.

\bibitem{lillicrap2015continuous}
T.~P. Lillicrap \emph{et~al.}, ``Continuous control with deep reinforcement learning,'' 2015, arXiv:1509.02971.

\bibitem{haarnoja2018soft}
T.~Haarnoja, A.~Zhou, P.~Abbeel, and S.~Levine, ``Soft actor-critic: Off-policy maximum entropy deep reinforcement learning with a stochastic actor,'' in \emph{Proceedings of the 35th International Conference on Machine Learning}, ser. Proceedings of Machine Learning Research, vol.~35, no.~1, Stockholm, 2018, pp. 1861--1870.

\bibitem{bemporad2007robust}
A.~Bemporad and M.~Morari, ``Robust model predictive control: A survey,'' in \emph{Robustness in identification and control}, ser. Lecture Notes in Control and Information Sciences, A.~Garulli and A.~Tesi, Eds.\hskip 1em plus 0.5em minus 0.4em\relax London, UK: Springer London, 2007, pp. 207--226.

\bibitem{mesbah2016stochastic}
A.~Mesbah, ``Stochastic model predictive control: An overview and perspectives for future research,'' \emph{IEEE Control Systems Magazine}, vol.~36, no.~6, pp. 30--44, 2016.

\bibitem{gros_data-driven_2020}
S.~Gros and M.~Zanon, ``Data-{Driven} {Economic} {NMPC} {Using} {Reinforcement} {Learning},'' \emph{IEEE Transactions on Automatic Control}, vol.~65, no.~2, pp. 636--648, 2020.

\bibitem{gros_learning_2022}
------, ``Learning for {MPC} with stability \& safety guarantees,'' \emph{Automatica}, vol. 146, p. 110598, 2022.

\bibitem{wachter2006implementation}
A.~W{\"a}chter and L.~T. Biegler, ``On the implementation of an interior-point filter line-search algorithm for large-scale nonlinear programming,'' \emph{Mathematical Programming}, vol. 106, pp. 25--57, 2006.

\bibitem{feldbaum1960dual}
A.~A. Feldbaum, ``Dual control theory. {I},'' \emph{Avtomatika i Telemekhanika}, vol.~21, no.~9, pp. 1240--1249, 1960.

\bibitem{wittenmark1995adaptive}
B.~Wittenmark, ``Adaptive dual control methods: An overview,'' \emph{IFAC Proceedings Volumes}, vol.~28, no.~13, pp. 67--72, 1995.

\bibitem{mesbah2018stochastic}
A.~Mesbah, ``Stochastic model predictive control with active uncertainty learning: A survey on dual control,'' \emph{Annual Reviews in Control}, vol.~45, pp. 107--117, 2018.

\bibitem{zanon2020safe}
M.~Zanon and S.~Gros, ``Safe reinforcement learning using robust {MPC},'' \emph{IEEE Transactions on Automatic Control}, vol.~66, no.~8, pp. 3638--3652, 2020.

\bibitem{lowrey2018plan}
K.~Lowrey, A.~Rajeswaran, S.~Kakade, E.~Todorov, and I.~Mordatch, ``Plan online, learn offline: Efficient learning and exploration via model-based control,'' 2019, arXiv:1811.01848.

\bibitem{lee2001NeurodynamicProgramming}
J.~M. Lee and J.~H. Lee, ``Neuro-dynamic programming method for {{MPC}},'' \emph{IFAC Proceedings Volumes}, vol.~34, no.~25, pp. 143--148, 2001.

\bibitem{lee2009ApproximateDynamic}
------, ``An approximate dynamic programming based approach to dual adaptive control,'' \emph{Journal of Process Control}, vol.~19, no.~5, pp. 859--864, 2009.

\bibitem{zhong2013Valuefunction}
M.~Zhong, M.~Johnson, Y.~Tassa, T.~Erez, and E.~Todorov, ``Value function approximation and model predictive control,'' in \emph{Proceedings of the {{IEEE Symposium}} on {{Adaptive Dynamic Programming}} and {{Reinforcement Learning}}}, 2013, pp. 100--107.

\bibitem{farshidian2019DeepValue}
F.~Farshidian, D.~Hoeller, and M.~Hutter, ``Deep value model predictive control,'' 2019, arXiv:1910.03358.

\bibitem{bhardwaj2020BlendingMPC}
M.~Bhardwaj, S.~Choudhury, and B.~Boots, ``Blending {{MPC}} \& value function approximation for efficient reinforcement learning,'' 2020, arXiv:2012.05909.

\bibitem{banker_gradient-based_2025}
T.~Banker and A.~Mesbah, ``Gradient-based framework for bilevel optimization of blackbox functions: Synergizing model-freereinforcementlearning and implicitfunctiondifferentiation,'' \emph{Industrial \& Engineering Chemistry Research}, vol.~64, no.~5, pp. 2831--2844, 2025.

\bibitem{chen_gnu-rl_2019}
B.~Chen, Z.~Cai, and M.~Bergés, ``Gnu-{RL}: Precocialreinforcementlearningsolution for buildinghvaccontrolusing a differentiable {MPC} policy,'' in \emph{Proceedings of the 6th {ACM} {International} {Conference} on {Systems} for {Energy}-{Efficient} {Buildings}, {Cities}, and {Transportation}}, New York, 2019, pp. 316--325.

\bibitem{amos_differentiable_2019}
B.~Amos, I.~D.~J. Rodriguez, J.~Sacks, B.~Boots, and J.~Z. Kolter, ``Differentiable {MPC} for end-to-end planning and control,'' 2019, arXiv:1810.13400.

\bibitem{romero2024ActorCriticModel}
A.~Romero, Y.~Song, and D.~Scaramuzza, ``Actor-critic model predictive control,'' 2024, arXiv:2306.09852.

\bibitem{hansenTemporalDifferenceLearning2022}
N.~Hansen, X.~Wang, and H.~Su, ``Temporal difference learning for model predictive control,'' 2022, arXiv:2203.04955.

\bibitem{anand_painless_2023}
A.~S. Anand, D.~Reinhardt, S.~Sawant, J.~T. Gravdahl, and S.~Gros, ``A painlessdeterministicpolicygradientmethod for learning-based {MPC},'' in \emph{Proceedings of the {European} {Control} {Conference}}, Bucharest, 2023, pp. 1--7.

\bibitem{mayne2000Constrainedmodel}
D.~Q. Mayne, J.~B. Rawlings, C.~V. Rao, and P.~O. Scokaert, ``Constrained model predictive control: {{Stability}} and optimality,'' vol.~36, no.~6, pp. 789--814, 2000.

\bibitem{maniaCertaintyEquivalenceEfficient2019}
H.~Mania, S.~Tu, and B.~Recht, ``Certainty equivalence is efficient for linear quadratic control,'' in \emph{Advances in Neural Information Processing Systems}, vol.~32, Vancouver, 2019.

\bibitem{gevers2005identification}
M.~Gevers, ``Identification for control: {{From}} the early achievements to the revival of experiment design,'' \emph{European Journal of Control}, vol.~11, no.~4, pp. 335--352, 2005.

\bibitem{klatt1998Gainschedulingtrajectory}
K.-U. Klatt and S.~Engell, ``Gain-scheduling trajectory control of a continuous stirred tank reactor,'' \emph{Computers \& Chemical Engineering}, vol.~22, no. 4--5, pp. 491--502, 1998.

\bibitem{fiedler2023DompcFAIR}
F.~Fiedler \emph{et~al.}, ``do-mpc: {{Towards FAIR}} nonlinear and robust model predictive control,'' \emph{Control Engineering Practice}, vol. 140, p. 105676, 2023.

\bibitem{bloor2024PCGymBenchmark}
M.~Bloor \emph{et~al.}, ``{{PC-Gym}}: Benchmark environments for process control problems,'' 2024, arXiv:2410.22093.

\bibitem{tan_sim--real_2018}
J.~Tan \emph{et~al.}, ``Sim-to-{Real}: Learningagilelocomotionforquadrupedrobots,'' 2018, arXiv:1804.10332.

\bibitem{witman2019sim}
M.~Witman, D.~Gidon, D.~B. Graves, B.~Smit, and A.~Mesbah, ``Sim-to-real transfer reinforcement learning for control of thermal effects of an atmospheric pressure plasma jet,'' \emph{Plasma Sources Science and Technology}, vol.~28, no.~9, p. 095019, 2019.

\bibitem{levine_offline_2020}
S.~Levine, A.~Kumar, G.~Tucker, and J.~Fu, ``Offline reinforcementlearning: Tutorial, review, and perspectives on openproblems,'' 2020, arXiv:2005.01643.

\bibitem{dontchev_implicit_2014}
A.~Dontchev and R.~Rockafellar, \emph{Implicit {Functions} and {Solution} {Mappings}: {A} {View} from {Variational} {Analysis}}, ser. Springer {Series} in {Operations} {Research} and {Financial} {Engineering}.\hskip 1em plus 0.5em minus 0.4em\relax New York: Springer, 2014.

\bibitem{rockafellar_directional_1984}
R.~T. Rockafellar, ``Directional differentiability of the optimal value function in a nonlinear programming problem,'' in \emph{Sensitivity, {Stability} and {Parametric} {Analysis}}, A.~V. Fiacco, Ed.\hskip 1em plus 0.5em minus 0.4em\relax Berlin, Heidelberg: Springer, 1984, pp. 213--226.

\bibitem{andrychowicz2017HindsightExperience}
M.~Andrychowicz \emph{et~al.}, ``Hindsight {{Experience Replay}},'' in \emph{Advances in Neural Information Processing Systems}, vol.~30, Long Beach, CA, 2017.

\bibitem{eysenbach2022contrastive}
B.~Eysenbach, T.~Zhang, S.~Levine, and R.~R. Salakhutdinov, ``Contrastive learning as goal-conditioned reinforcement learning,'' in \emph{Advances in Neural Information Processing Systems}, vol.~35, New Orleans, 2022, pp. 35\,603--35\,620.

\bibitem{salzmann2024LearningCasADi}
T.~Salzmann, J.~Arrizabalaga, J.~Andersson, M.~Pavone, and M.~Ryll, ``Learning for {CasADi}: {D}ata-driven models in numerical optimization,'' in \emph{Proceedings of the 6th Annual Learning for Dynamics \& Control Conference}, ser. Proceedings of Machine Learning Research, vol. 242, Oxford, 2024, pp. 541--553.

\bibitem{schulman_high-dimensional_2018}
``High-{Dimensional} continuouscontrolusinggeneralizedadvantage estimation.''

\bibitem{kumar_stabilizing_2019}
A.~Kumar, J.~Fu, M.~Soh, G.~Tucker, and S.~Levine, ``Stabilizing offpolicy {Q}-learning via bootstrappingerrorreduction,'' in \emph{Advances in Neural Information Processing Systems}, vol.~32, Vancouver, 2019.

\bibitem{andersson2019casadi}
J.~A. Andersson, J.~Gillis, G.~Horn, J.~B. Rawlings, and M.~Diehl, ``{CasADi}: a software framework for nonlinear optimization and optimal control,'' \emph{Mathematical Programming Computation}, vol.~11, pp. 1--36, 2019.

\bibitem{janner2019WhenTrust}
M.~Janner, J.~Fu, M.~Zhang, and S.~Levine, ``{When} to trust your model: Model-based policy optimization,'' in \emph{Advances in Neural Information Processing Systems}, vol.~32, Vancouver, 2019.

\end{thebibliography}

\end{document}